\documentclass[10pt, twocolumn,letterpaper]{article}
\nonstopmode

\usepackage{wacv}
\usepackage{times}
\usepackage{epsfig}
\usepackage{graphicx}
\usepackage{amsmath}
\usepackage{amssymb}
\graphicspath{{images/}}
\DeclareGraphicsExtensions{.png,.jpg,.pdf}

\usepackage{graphicx}
\usepackage{booktabs} 
\usepackage{subcaption}
\usepackage{array}
\usepackage{tabularx}
\newcolumntype{L}[1]{>{\raggedright\arraybackslash}p{#1}}
\newcolumntype{K}[1]{>{\centering\arraybackslash}p{#1}}
\usepackage{color}
\usepackage[ruled]{algorithm2e} 

\wacvfinalcopy 

\def\wacvPaperID{0318} 

\ifwacvfinal\fi
\begin{document}


\newcommand{\Var}{\mathrm{Var}}

\DeclareRobustCommand{\rchi}{{\mathpalette\irchi\relax}}
\newcommand{\irchi}[2]{\raisebox{\depth}{$#1\chi$}} 

\newcommand{\trimmedteaser}[1]{\frame{\includegraphics[trim=6cm 3cm 6cm 2cm,clip,width=1.75in]{#1}}}

\newcommand*{\linkback}[1]{\Acrobatmenu{GoBack}{\color{black}#1}}

\newcommand{\linkoverview}[1]{\hyperlink{page.\getpagerefnumber{lbl:overview}}{#1}}

\newcommand*\rot{\rotatebox{90}}
\newcommand*\chechmark{\ding{51}}
\newcommand*\OK{\ding{51}}

\newcommand{\jumplabel}[1]{\phantomsection\label{#1}}

\newcommand{\citem}{\vspace{-1.5ex}\item}
\newcommand{\citemarg}[1]{\vspace{-1.5ex}\item[\changed{#1}]}

\newlength{\w}
\newlength{\h}
\newlength{\x}

\newcommand{\cfbox}[2]{%
    \colorlet{currentcolor}{.}%
    {\color{#1}%
    \fbox{\color{currentcolor}#2}}%
}

\newcommand{\indenttbl}[0]{\hspace{0.2cm}}
\newcommand{\pmpad}[0]{\hspace{0.09cm}\pm\hspace{0.09cm}}
\newcommand{\ignorethis} [1] {}

\newcommand{\sectnum    } [1] {\ref{#1}}
\newcommand{\sect       } [1] {Section~\sectnum{#1}}
\newcommand{\sects      } [1] {Sections~\sectnum{#1}}
\newcommand{\shortsect  } [1] {\S\sectnum{#1}}

\newcommand{\tbl}[1]{Table~\ref{#1}}

\newcommand{\fignum     } [1] {\ref{#1}}
\newcommand{\fig        } [1] {Figure~\fignum{#1}}

\newcommand{\eqnnum     } [1] {\mbox{(\ref{#1})}}
\newcommand{\eqn        } [1] {equation~\eqnnum{#1}}

\newcommand{\X}{$\times$\xspace}
\newcommand{\tX}{\hspace{-0.1em}$\small\times$\hspace{-0.2em}}

\newcommand{\Mod}[1]{\ (\text{mod}\ #1)}

\newcommand{\shrinkbefore}{\vspace{-0.6em}}
\newcommand{\shrinkafter}{\vspace{-1.4em}}

\definecolor{verydarkgreen}{rgb}{0,0.2,0}
\definecolor{verydarkorange}{rgb}{0.4,0.2,0}
\definecolor{verydarkyellow}{rgb}{0.55,0.55,0}

\newcommand{\changedshort}[1]{{#1}}
\newcommand{\changed}[1]{{#1}}
\newcommand{\changedb}[1]{{#1}}
\newcommand{\changedc}[1]{{#1}}
\newcommand{\changedd}[1]{{#1}}

\ifdefined\ShowNotes
\newcommand{\Connelly}[1]{\colornote{darkgreen}{Connelly}{#1}}
\newcommand{\Fuwen}[1]{\colornote{dblue}{Fuwen}{#1}}
\newcommand{\Yuting}[1]{\colornote{maroon}{Yuting}{#1}}
\newcommand{\todo}[1]{\colornote{darkgreen}{Todo}{#1}}
\else
\newcommand{\Connelly}[1]{}
\newcommand{\Quanquan}[1]{}
\newcommand{\Sam}[1]{}
\newcommand{\Yuting}[1]{}
\newcommand{\todo}[1]{}
\fi

\newcommand{\change}[1]{{#1}}
\newcommand{\changeb}[1]{{#1}}

\newcommand{\dxy}[0]{$g_{xy}$}
\newcommand{\dwh}[0]{$g_{wh}$}

\newcommand{\videoframerate}[0]{1}

\newcommand{\treecannspeedup}[0]{12$\times$}
\newcommand{\kcoherencespeedup}[0]{9$\times$}
\newcommand{\patchmatchspeedup}[0]{200$\times$}

\ifdefined\ShowNotes
  \newcommand{\colornote}[3]{{\color{#1}\bf{#2: #3}\normalfont}}
\else
  \newcommand{\colornote}[3]{}
\fi

\definecolor{darkred}{rgb}{0.7,0.1,0.1}
\definecolor{darkgreen}{rgb}{0.1,0.7,0.1}
\definecolor{verydarkgreen}{rgb}{0.0,0.5,0.0}
\definecolor{verydarkblue}{rgb}{0.0,0.0,0.7}
\definecolor{cyan}{rgb}{0.7,0.0,0.7}
\definecolor{dblue}{rgb}{0.2,0.2,0.8}
\definecolor{maroon}{rgb}{0.76,.13,.28}
\definecolor{burntorange}{rgb}{0.81,.33,0}

\newcommand{\INDSTATE}[1][1]{\STATE\hspace{#1\algorithmicindent}}

\newcommand\fnurl[2]{%
  \href{#2}{#1}\footnote{\url{#2}}%
}


\newcommand\ulinec{\bgroup\markoverwith
      {\textcolor{uhref_color}{\rule[-0.5ex]{2pt}{0.4pt}}}\ULon}

\newcommand{\nsfurl}[1]{\ulinec{\small{\color{uhref_color}{\url{#1}}}}}

\newcommand{\beginintegraltable}[2]{
\begin{tabular}{|l|l|}
	\hline
	Function $f(x)$ & Bandlimited with #1 kernel: $\hat{f}^{(#2)}(x)$ \\
	\hline
}
\newcommand{\finishintegraltable}{
	\hline
	\end{tabular}\\ \\
}

\title{Where and Who? Automatic Semantic-Aware Person Composition}

\author{Fuwen Tan \\
University of Virginia\\
\and
Crispin Bernier \\
University of Virginia\\
\and
Benjamin Cohen \\
University of Virginia\\
\and
Vicente Ordonez \\
University of Virginia\\
\and
Connelly Barnes \\
University of Virginia\\
}

\maketitle
\ifwacvfinal\thispagestyle{empty}\fi

\begin{abstract}
   Image compositing is a method used to generate realistic yet fake imagery by inserting contents from one image to another. Previous work in compositing has focused on improving appearance compatibility of a user selected foreground segment and a background image (i.e. color and illumination consistency).
   In this work, we instead develop a fully automated compositing model that additionally learns to select and transform compatible foreground segments from a large collection given only an input image background.
   To simplify the task, we restrict our problem by focusing on human instance composition, because human segments exhibit strong correlations with their background and because of the availability of large annotated data.
   We develop a novel branching Convolutional Neural Network (CNN) that jointly predicts candidate person locations given a background image. We then use pre-trained deep feature representations to retrieve person instances from a large segment database. Experimental results show that our model can generate composite images that look visually convincing.
   We also develop a user interface to demonstrate the potential application of our method.
\end{abstract}


\section{Introduction}

\label{sec:intro}

Image compositing aims to produce images that can trick humans into believing they are real, although they are not. Image composites can also result in fantastic images that are limited only by an artist's imagination. However, the process of creating composite images is challenging, and it is not fully understood how to make realistic composites. A typical compositing task proceeds in four steps:
(1) choose a foreground segment that is semantically compatible with a given background scene;
(2) place the segment at a proper location with the right size;
(3) perform operations such as alpha matting~\cite{Smith1996} or Poisson blending~\cite{Perez2003} to adjust the local appearance;
(4) apply global refinements such as relighting or harmonization~\cite{Tsai2017}.
The first two steps require semantic reasoning while the last two steps deal with appearance compatibility.
Whether a human perceives a composite image as real or fake depends on all these factors.
However, while existing compositing systems tackle the last two steps automatically, most of them leave the semantic tasks (steps 1 and 2) to the users.

In this work, we explore the semantic relationships between a collection of foreground segments and background scenes using a data-driven method for automatic composition. 
We restrict the foreground category to ``human" because humans play a central role in a large proportion of image composites, and because we can easily collect enough exemplar data for training and testing. For simplicity, we also choose to ignore occlusion by assuming that human segments are fully visible from the camera viewpoint. 

This research is motivated by recent breakthroughs in scene recognition~\cite{Xiao2016} and object-level reasoning~\cite{Ren2015} through deep neural networks, which have brought unprecedented levels of performance for similar semantic tasks. Thus, we apply these techniques to estimate the semantic compatibility between candidate foreground segments and image backgrounds using a large scale visual dataset. Given these observations, our method contains three components: First, a location proposal step where we predict the location and size of each potential person instance using a novel Convolutional Neural Network (CNN) architecture. Second, a retrieval step, where we find a specific segment that semantically matches the local and global context of the scene. Lastly, a final compositing step, where we leverage off-the-shelf alpha matting~\cite{Chen2012} to adjust the transition between a composited segment and its surroundings so that the segment appears compatible with the background.




To evaluate our method, we conduct quantitative and qualitative experiments including a user study. 
We demonstrate that our generation pipeline can be useful for interactive layout design or storyboarding tasks which can not be easily fulfilled using other tools. 

We summarize here our technical contributions: 
(1) A model that predicts probable locations for the presence of a person instance for an arbitrary input background using contextual cues.
(3) A fully automatic person compositing system which generates convincing composite images. To the best of our knowledge, this is the first attempt towards this task;
(4) Conducting quantitative and qualitative evaluations, including a user study and a proof-of-concept user interface.

\section{Related work}

\textbf{Composite image generation.} Early methods for compositing such as alpha matting~\cite{Smith1996} and gradient-domain compositing~\cite{Perez2003} can seamlessly stitch a foreground object with a background image by blending a local transition region. To enforce global appearance compatibility, Lalonde and Efros \cite{lalonde2007} proposed to model the co-occurrence probability of the foreground object and the background image using a color distribution.
Similarly, Xue~et~al.~\cite{Xue2012} proposed to investigate the key statistical properties that control the realism of an image composite.
Recently, Zhu~et~al.~\cite{zhu2015} trained a single CNN-based model to distinguish composite images from natural photographs and refine them by optimizing the predicted scores.
Furthermore, Tsai~et~al.~\cite{Tsai2017} developed an end-to-end deep CNN based model for image harmonization.
These methods give visually pleasing results, but unlike our work, they all leave the semantic tasks to the users, such as choosing foreground segments and placing them at proper positions with the right size.

The work of Lalonde~et~al.~\cite{lalonde07b} took a step further by building an interactive system to insert new objects into existing photographs by querying a vast image-based object library. Chen~et~al.~\cite{Chen2009} developed a similar interactive system but took user sketches as input.
Hays~et~al.~\cite{Hays2007} proposed an automatic patch retrieval and blending method for scene completion using millions of photographs. Unlike our work, these methods relied on hand-crafted features and the composite regions were still indicated manually by the users.

\textbf{Context based scene reasoning.} 
Using context for scene reasoning has a long history~\cite{Divvala2009}. 
Pioneering works include Bar and Ullman~\cite{Strat1991} and Strat and Fischler~\cite{Moshe1996}, which incorporated contextual information for recognition. 
Context based methods are also popular in object-level classification.
Bell~et~al.~\cite{bell2016} proposed a Recurrent Neural Network framework to detect objects in context.
These works modeled correlations among contents within the image, while our method predicts contents that are not yet present.
Related to our work, Torralba~et~al.~\cite{Torralba2003a} introduced a  challenge to test to what extent can object detection succeed by only contextual cues. 
More recently, Sun~et~al.~\cite{sun2017} proposed a siamese network to detect missing objects in an image.
While these methods predicted contents that were not present in the images, they all focused on the binary determination of whether there should be any object at a specific location or not.
In contrast, our method attempts to predict both the location and size of a potential foreground segment, and retrieve a segment with proper appearance that is compatible with the surrounding context. 
Concurrently to our research, Wang et al.~\cite{wang2017} proposed to model affordances by predicting the skeletons of persons that were not already present. 
This method achieves good results, but requires matching of a similar indoor scene, and only predicts a skeleton, not a full color composite. Finally, Kermani~et~al.~\cite{kermani2016learning} synthesized 3D scenes by learning factor graph and arrangement models from an RGB-D dataset.

\begin{figure*}[t]
  \centering
  \includegraphics[width=0.95\textwidth]{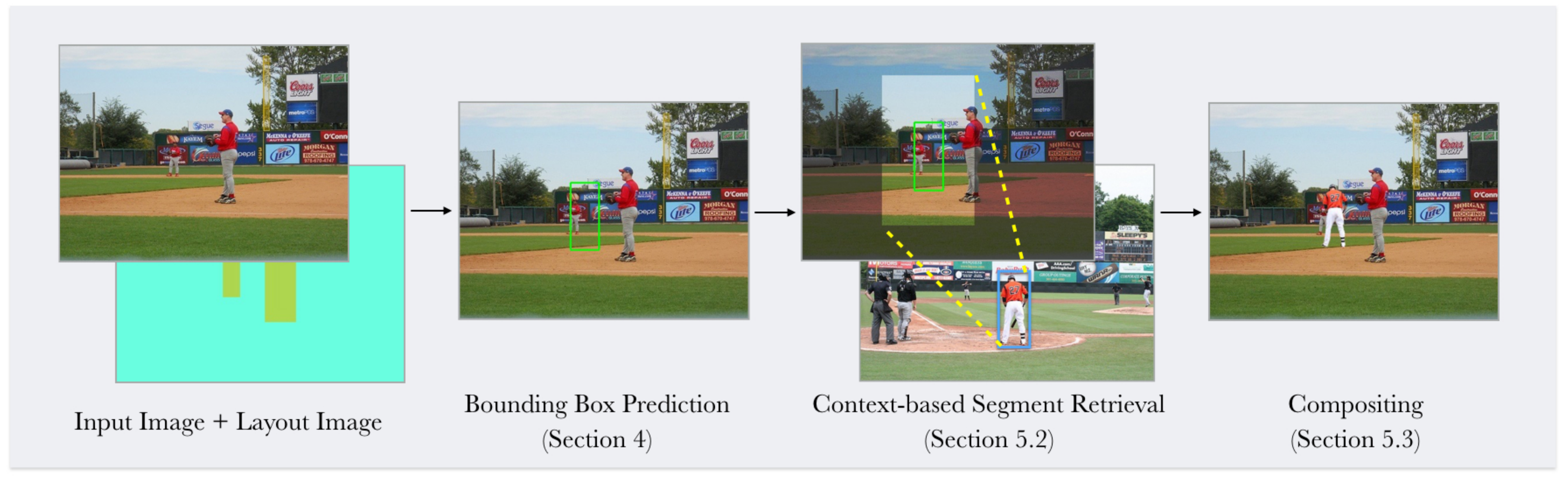}
  \vspace*{-3.5mm}
  \caption{Overview of our pipeline: our system consists of three computational stages, which are depicted above. }
  \label{fig:overview}
  \vspace*{-2mm}
\end{figure*}

\textbf{Context based image editing.} 
By conditioning on the local surroundings, Pathak~et~al.~\cite{pathak2016} performed semantic inpainting by using a generative adversarial network. 
Yang~et~al.~\cite{yang2017} proposed a multi-scale CNN model for high-resolution image inpainting via neural patch synthesis.
Iizuka~et~al.~\cite{Iizuka2017} developed a CNN based method for image completion by enforcing global and local consistency.
Recently, Chen~et~al.~\cite{chen2017} presented a cascaded refinement network to synthesize images conditioned on semantic layouts.
While these methods synthesize novel contents from context at pixel level, the locations or layouts of the synthesized regions were still provided by the users. 
Our method predicts such regions and retrieves plausible segments.

\section{Overview}
Figure~\ref{fig:overview} shows an overview of our system. It has three main components: bounding box prediction, person segment retrieval, and compositing. We now give a brief discussion of each of these.

In \sect{sec:boxprediction}, we introduce our proposed CNN based model to predict a bounding box of the potential segment.
We formulate the bounding box prediction as a joint classification problem by discretizing the spatial and size domains of the bounding box space. 
Specifically, we design a novel two branch network which can be trained end-to-end using supervised learning, and tested in a cascade manner. 

In \sect{sec:retrieval}, we introduce a candidate pool we built for segment retrieval. 
A context based segment retrieval scheme is devised to find a person segment from the candidate pool that semantically matches both local context and the global scene.
The key component for achieving this is a hybrid deep feature representation.
Finally, we use an alpha matting technique to composite the retrieved segment with the background at the predicted location and size.

In \sect{sec:evaluation}, we evaluate our bounding box prediction model quantitatively by measuring the histogram correlation between the ground truth bounding boxes distribution and our prediction. 
We also evaluate the visual realism of composite images with a human subject evaluation.




\section{Bounding box prediction}
\label{sec:boxprediction}
In this section, we introduce our learning based method to predict the bounding box of a potential person segment given a single background image.
Our key insight here is that the correlation between the foreground segment and the background scene can be learned directly from human-annotated object layouts of natural images.

We first discuss how we collect and preprocess the data (\sect{sec:prediction_preprocessing}). Next, we explain the input for the  model (\sect{sec:prediction_input}), and give the prediction target for the model (\sect{sec:prediction_target}). We then give the model itself (\sect{sec:prediction_model}).

\subsection{Data preprocessing}
\label{sec:prediction_preprocessing}
The data we use for learning such layout correlation is from  MS-COCO~\cite{Lin2014}. This dataset contains tens of thousands of images with both bounding box and segment annotations for each object instance in 80 different categories.

Because a large proportion of object instances are occluded, we automatically filter out heavily occluded person instances using three passes of filtering: 
(1) We filter the person instances whose bounding boxes
have large overlapping areas with other objects. 
Specifically, we exclude instances whose Intersection over Union (IoU) with any other instance is larger than 0.3. 
(2) We also exclude person instances that are close to the edge of the images as they are probably incomplete. 
In particular, we filter the instance if the distance between its bounding box and the edge of the image is less than 18 pixels.
(3) Finally, we remove instances whose areas are less than 2500 square pixels. 

After applying the filtering routines, we obtain 36,636 person instances from the training split of MS-COCO, and 16,962 from the validation split.

\subsection{Input imagery}
\label{sec:prediction_input}


For each person instance in the dataset, we attempt to learn the mapping from its background context to the person's bounding box.
Learning such a mapping function requires us having an input image in which the person is not already present.
However, to do this, we have to "erase" the person instances from the source images.
Our solution is to remove the person instances automatically by using the human-annotated segments from MS-COCO.
We remove each person via the inpainting method of Barnes~et~al.~\cite{barnes2009patchmatch}, implemented as Content Aware Fill from Adobe Photoshop. 
The resulting inpainted results sometimes exhibit artifacts such as repetitive patches. 
To prevent the model from over-fitting on these artifacts, the inpainted image is further blurred using a Gaussian with a sigma of 3.2.
We denote the blurred image as $I_B$.

Given the recent breakthroughs in CNN-based object detection systems~\cite{Ren2015,bell2016}, in addition to using our inpainted (and blurred) images directly as input, we also incorporate the informative output from an object detector.
We use the Faster RCNN object detector~\cite{Ren2015} to obtain object detections in the inpainted images.
The bounding boxes of the detected objects in different categories are then rendered using a randomly generated color palette, with each color corresponding to a category. 
The color values within an overlapping region are set to the mean color value.
We find that using different color palettes achieves similar performance. The layout image (indicated as $I_L$) represents the object layout of the image, as shown in \fig{fig:overview}.


\subsection{Prediction target}
\label{sec:prediction_target}
The target of our prediction model is the bounding box of a person. We first discuss how we represent the bounding box using normalized coordinates, and then explain how we discretize these coordinates for use in classification.

The bounding box representation from a ground truth annotation in the dataset is a four dimensional vector: ($x_\mathrm{min}$, $y_\mathrm{min}$, $x_\mathrm{max}$, $y_\mathrm{max}$), where ($x_\mathrm{min}$, $y_\mathrm{min}$) represents the top-left coordinate and ($x_\mathrm{max}$, $y_\mathrm{max}$) represents the bottom-right coordinate.
For images of different resolutions, a normalized bounding box representation is required for consistent prediction.
To do this, our system pads each rectangular image by the minimum amount so a square image is obtained, using a padding color that is the mean color for the ImageNet dataset~\cite{ILSVRC15}. 
The bounding box is first shifted to account for the square padding, then transformed into normalized coordinates $(x_\mathrm{stand}, y_\mathrm{stand}, w, h) \in [0, 1]$, where $x_\mathrm{stand} = \frac{1}{2s}(x_\mathrm{min} + x_\mathrm{max})$, $y_\mathrm{stand} = \frac{1}{s}y_\mathrm{max}$, $s$ is the width of the square image, and $w$, $h$ are the width and height of the box relative to the square image.
Thus, $(x_\mathrm{stand}, y_\mathrm{stand})$ is the lowest center (standing) point of the bounding box.

Direct regression in a four dimensional continuous space is challenging.
To facilitate the bounding box prediction, our system discretizes the $(x, y)$ location domain into a $15 \times 15$ grid board, and then represents $(x_\mathrm{stand}, y_\mathrm{stand})$ as the index of the grid \dxy{} where it is located.
Similarly, the $(w, h)$ size domain is also discretized so that $(w, h)$ is  represented as another grid index \dwh{}.
By doing this, we formulate the bounding box prediction as two classification problems with 225 (15 x 15) different classes for each.

\subsection{Prediction model}
\label{sec:prediction_model}

\begin{figure}[t]
  \centering
  \includegraphics[width=0.48\textwidth]{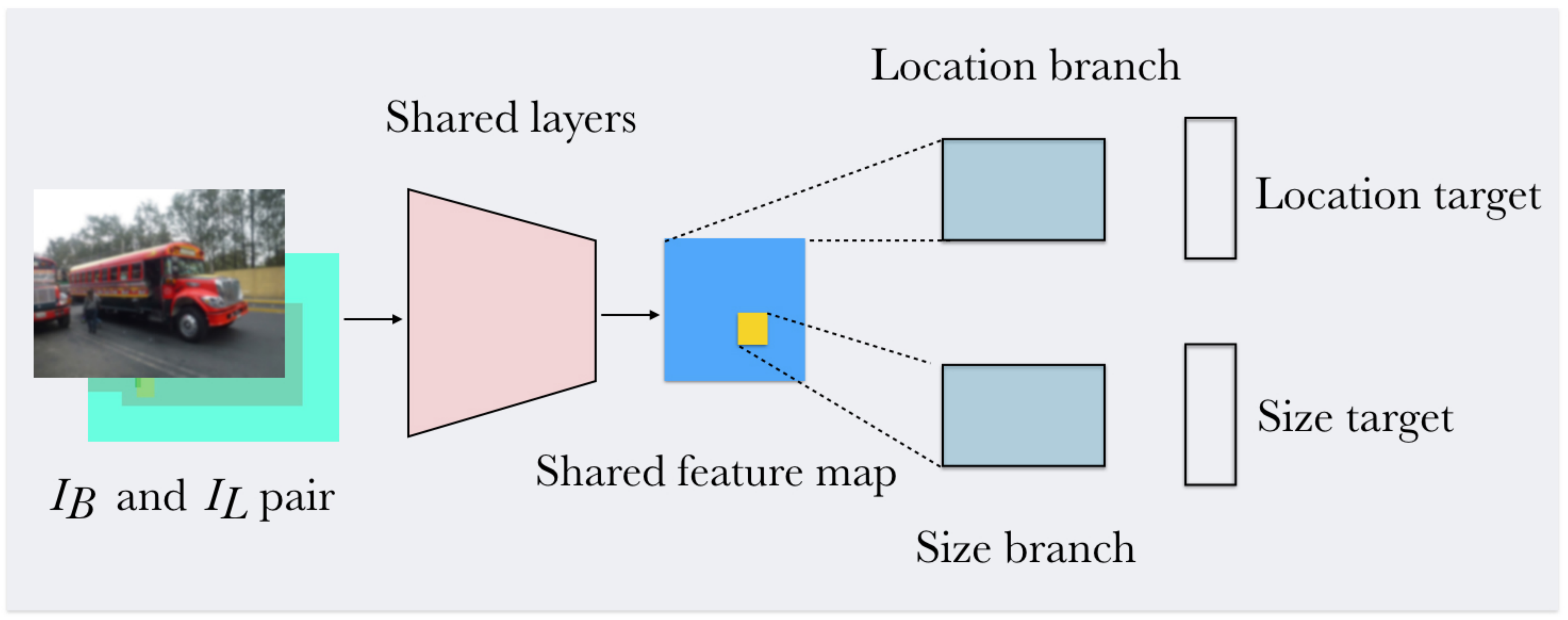}
  \vspace*{-7mm}
  \caption{Overview of the prediction model: the proposed network exploits a two branch architecture: the first branch predicts location and the second branch predicts size.}
  \label{fig:prediction}
  \vspace*{-5mm}
\end{figure}

Given $I_B$, $I_L$ as inputs and \dxy{}, \dwh{} as targets, our next challenge is to learn the underlying mapping between them. 
Our approach is to learn the location (\dxy{}) and size (\dwh{}) simultaneously as they are highly correlated.
In particular, we develop a novel CNN-based model which can be trained in an end-to-end manner.

In our model, the images $(I_B, I_L)$ are first concatenated along the depth channel, and fed through a shared front-end network, as shown in \fig{fig:prediction}. This network is shared in the sense that the same weights are used before the split into the location and size branches.
The shared network contains three residual bottleneck modules with projection shortcuts, similarly as in He~et~al.~\cite{He2016}. 
Starting from the output feature map of the shared network, the model is then separated into two smaller branches, with the first branch predicting the location (\dxy{}), and the second branch predicting the size (\dwh{}).
These two branches also incorporate dilated convolutional layers introduced in~\cite{Yu2016} in order to use larger receptive fields without using an additional number of parameters.
Table 1 lists the layer-by-layer details of the proposed network architecture.

\begin{table}
    \small\selectfont
    \centering
    \label{tab:arch}
    \vspace*{1mm}
    \begin{tabular}{L{5.0cm}L{2.5cm}}
        \hline Layers&Activation Size \\
        \hline
        \multicolumn{1}{l}{\textit{Shared layers}}\\
        \indenttbl{}Input&6 x 480 x 480 \\
        \indenttbl{}Conv: 64 x 7 x 7, stride 2 & 64 x 237 x 237 \\
        \indenttbl{}Max pool 3 x 3, stride 2 & 64 x 118 x 118 \\
        \indenttbl{}Conv block, (64, 64, 128) filters* & 128 x 59 x 59 \\
        \indenttbl{}Conv block, (64, 64, 128) filters* & 128 x 30 x 30 \\
        \indenttbl{}Conv block, (128, 128, 512) filters* & 512 x 15 x 15 \\
        \hline 
        \multicolumn{1}{l}{\textit{Location prediction branch}}\\
        \indenttbl{}Conv: 64 x 3 x 3, dilation 2 & 64 x 15 x 15 \\
        \indenttbl{}Conv: 1 x 3 x 3, dilation 2 & 15 x 15 \\
        \hline 
        \multicolumn{1}{l}{\textit{Size prediction branch}}\\
        \indenttbl{}Conv: 512 x 3 x 3, dilation 2 & 512 x 15 x 15 \\
        \indenttbl{}ROI slicing & 512 x 3 x 3 \\
        \indenttbl{}Global maxpooling & 512 \\
        \indenttbl{}Two fully connected layers & 225 \\
        \hline 
    \end{tabular}\\
    \vspace{-3mm}
    \caption*{Table 1. Architecture of our prediction model.\\ * The last layer has stride 2 and a projection shortcut~\cite{He2016}.}
    \vspace*{-5mm}
\end{table}

Note that the size of the predicted box should be consistent with the local context. 
For instance, person segments should not be larger than instances of larger objects appearing in their surroundings, such as buses or cars. 
Therefore, in the size prediction branch, after a small 3 x 3 dilated convolution, 
our system first remaps the normalized coordinates $(x_\mathrm{stand}, y_\mathrm{stand})$ into the spatial coordinates of the output activations, to obtain grid coordinates $(x_\mathrm{grid}, y_\mathrm{grid})$.
A  (512 x 3 x 3) activation slice is then extracted along the depth channel. This is done by extracting activations from a box with 3x3 spatial size, such that the lowest center coordinate of the box is $(x_\mathrm{grid}, y_\mathrm{grid})$. 
We call this process Region of Interest (ROI) slicing.
This smaller activation map is then fed through the rest of the layers of the size branch.
By doing this, the size prediction network attends to a sub-region of the feature map that captures the local context.

One subtle point for this design is that, during training, the normalized coordinates $(x_\mathrm{stand}, y_\mathrm{stand})$ we use for ROI slicing come from the ground truth bounding box.
However, during testing, $(x_\mathrm{stand}, y_\mathrm{stand})$ are generated from the location we predict.
Therefore, during inference our network runs in two stages: the first stage predicts only location; the second stage predicts the size based on the location. Please see our supplemental for additional implementation details.



\section{Person segment retrieval and compositing.}
\label{sec:retrieval}

In this section, we introduce a simple context-based person segment retrieval and compositing scheme based on a hybrid deep feature representation. We first discuss how we create the pool of candidate person segments and perform  retrieval. Then we describe how we perform compositing.

\subsection{Creating the candidate pool of person segments}
\label{sec:candidate_pool}
To build a candidate pool for person segment retrieval, we use the annotated data from the validation split of the MS-COCO dataset. We chose this split because these images are also held out from the training of the bounding box prediction.
We apply the same filtering routines as in Section 4.1 to exclude segments that are heavily occluded, small or incomplete.
Finally, we manually filter the remaining segments to remove partially occluded instances. 
In total, our candidate pool contains 4100 person segments.

Although these segments come with ground truth segmentation annotations, most of the annotations are not accurate enough for compositing applications.
Therefore, we also perform manual segmentation using the lasso tool from Adobe Photoshop for the segments we present and used in the human subject study (see Section 6).

To demonstrate the generalization of our system, all the background testing images we present in the paper are from the YFCC100M split of the VisualGenome dataset~\cite{Krishna2017} and the SUN dataset~\cite{Xiao2010}.

\subsection{Context based person segment retrieval}

\begin{figure}[t]
  \centering
  \includegraphics[width=0.48\textwidth]{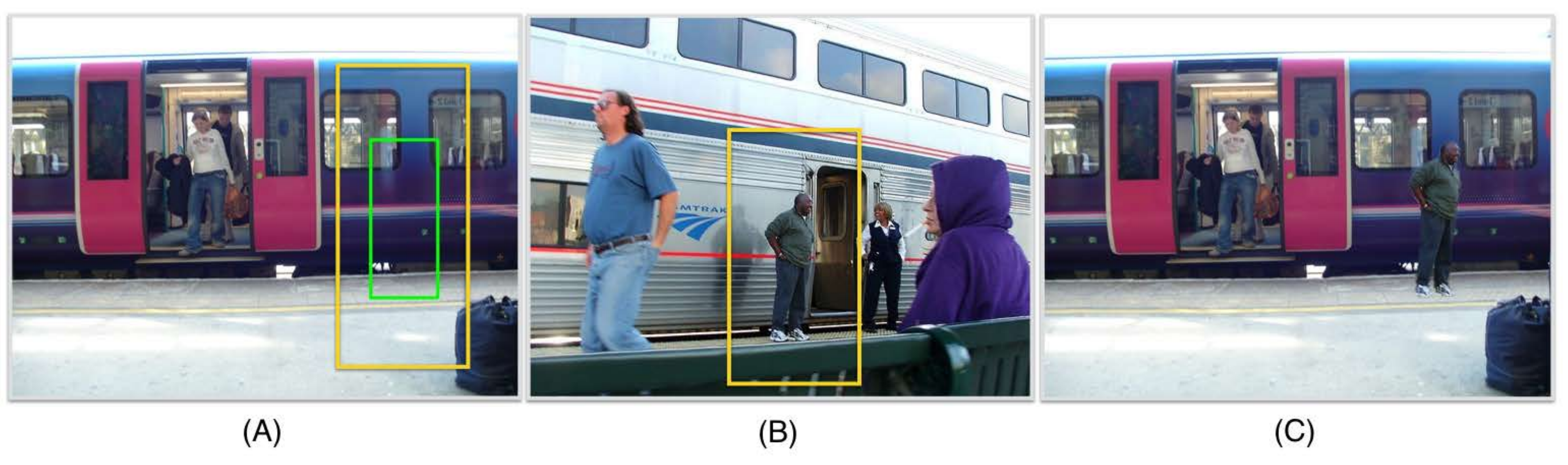}
  \vspace*{-8mm}
  \caption{Person segment retrieval: given the input image (A) and the predicted bounding box (the green box in (A)), the proposed system incorporates features from both the global scene and the local context (covered by the yellow box in (A)) to retrieve a favorable person segment (within the yellow box in (B)) and composite it on the input image.}
  \label{fig:retrieval}
  \vspace*{-5mm}
\end{figure}

\begin{figure*}[t]
  \centering
  \includegraphics[width=1.0\textwidth]{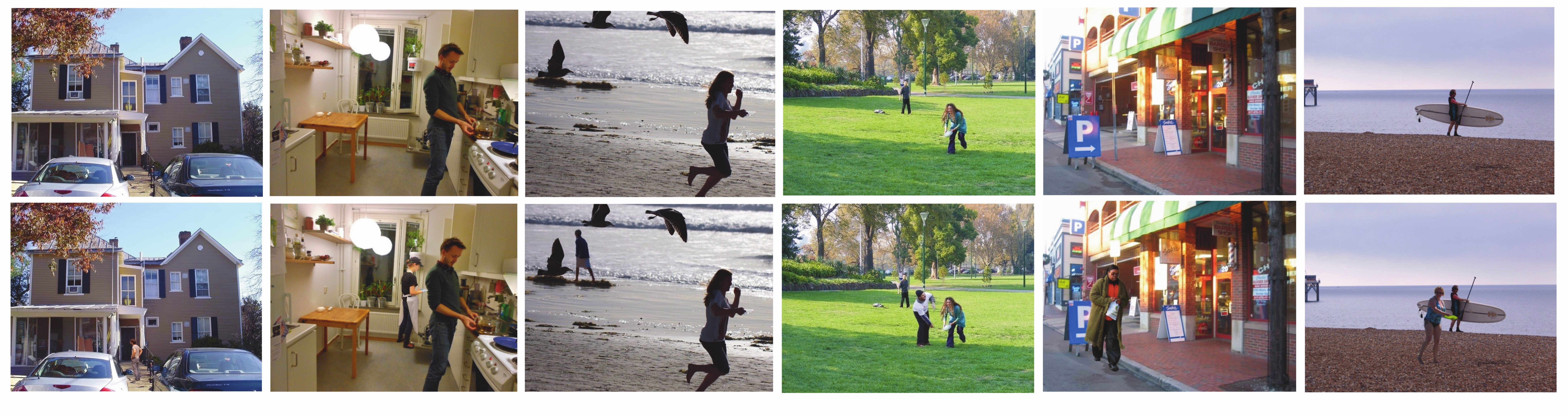}
  \vspace*{-9mm}
  \caption{Composites automatically generated from our system. The first row shows the input images, the second row shows the composite results. We include additional results in the supplemental.}
  \label{fig:results}
  \vspace*{-5mm}
\end{figure*}

Given a background image and a predicted bounding box,
our goal is to retrieve a person segment from the candidate pool that not only matches the global scene semantics but also appears compatible with the local context. Various hand-crafted feature descriptors \cite{Torralba2003b}\cite{Hu2013} have been proposed to facilitate image retrieval.
Recently, using intermediate neural network activations as feature representations has shown to perform competitively for various semantic retrieval tasks even when the underlying network has been pre-trained in an unrelated classification task~\cite{Babenko2014}.
However, previous methods mostly aim to retrieve images that ``look similar'' with respect to a query image, while our goal is to retrieve segments which are not present but ``look natural'' when composited on a background scene.

Our key insight here is that, by incorporating the contextual information of both the query background image and the candidate person segments, we could adapt and extend feature-based methods to retrieve segments from images which share similar global scene semantics and local context with the background image. Specifically, for each input image, our system first extracts deep features which describe global scene semantics of the background image.
We adopt the activation map from the mean pooling layer of ResNet50~\cite{He2016}.
Similarly, for each candidate person segment, we extract the same feature descriptor for its background image.
Measuring the distance between the input image and the candidate images in feature space can help retrieve segments appearing in similar scenes.
However, the retrieved segment does not necessarily look natural in the local context if only global compatibility is considered.

To further enforce the local compatibility, given the predicted bounding box, our system crops a local image patch which shares the same center with the bounding box but is twice as large in both width and height, as shown in \fig{fig:retrieval}.
The same feature descriptor (activations of the mean pooling layer of ResNet50) of this local patch is then extracted.
For each candidate person segment, our system extracts similar local feature descriptors.
Measuring the distance between these local features can help retrieve segments appearing in similar local contexts as in the target location. 

In our implementation, the segment retrieval proceeds in two steps: (1) Our system first filters the segments whose bounding box sizes are quite different from the query box size.
To do this, our system aligns the centers of the query and target bounding boxes and computes their Intersection over Union (IoU). Segments with IoUs smaller than 0.4 are excluded.
(2) From the remaining candidate segments in our collection, the system retrieves the top one segment that is ``closest" to the query input in feature space.
Specifically, we use cosine distance between the query and the target segment, each represented by a concatenation of the global and local feature descriptors.
To accelerate the retrieval process, we also build a kd-tree structure of the candidate segments.

\subsubsection{Selection of features}
For the retrieval task, we experimented with a few different feature descriptors (e.g. GIST feature~\cite{Torralba2003b}, unsupervised learned feature from the Context Encoder work ~\cite{pathak2016}, deep activation maps from VGG16~\cite{Simonyan14c} and ResNet50). We adopted the mean pooling feature from ResNet50 based on several observations: 
(1) Compared with GIST and context encoder features~\cite{pathak2016},
the superiority of deep features was demonstrated by a pilot user study we conducted on Amazon Mechanical Turk. The setup of the pilot study was similar with the one we are going to introduce in~\sect{sec:user_study}. 
(2) Different from VGG16, ResNet50 incorporated Batch Normalization layers~\cite{Ioffe2015}, which produce activation maps with similar magnitude scales for different dimensions. 
This is important when measuring the distance between features. 
Because the feature maps from VGG16 exhibited various magnitude scales for different dimensions, our experiments showed that they usually resulted in poor retrieved segments.
(3) Compared with other layers in ResNet50, the activation map from the mean pooling layer encodes much semantic information (it is one layer before the final classifier) in smaller dimensions (2048), which makes it both effective and efficient for our retrieval task. 

\subsection{Compositing}
With the retrieved segment in hand, our system scales and composites it onto the background such that the segment has the same center and height as the predicted bounding box.
Although the segment already has a clean binary mask produced from the Photoshop magnetic lasso tool we discussed in \sect{sec:candidate_pool}, we apply an off-the-shelf alpha matting method ~\cite{Chen2012} to obtain smooth natural transitions over the composite region. 
\fig{fig:results} shows example composites produced by our method covering various scenes. 
As our current pipeline has not considered relighting explicitly, the composites may suffer from lighting inconsistency problems. We  leave relighting to future work.

\section{Evaluation}
\label{sec:evaluation}
\subsection{Quantitative evaluation of box prediction}
During training of the bounding box prediction model, we use the ground truth bounding boxes as the target for supervised learning.
At first glance, it may seem reasonable to use evaluation metrics from object detection systems, such as average precision or precision-recall (PR) curve.
However, for each specific background image, there may be multiple locations suitable for composing person instances with various sizes.
The goal of the prediction model is to learn the distribution of feasible object layouts instead of overfitting toward the exact ground truth boxes in the dataset.
In fact, we try to avoid this situation by blurring the input image so that the system can not overfit to inpainting artifacts.

Therefore, to evaluate the performance of the bounding box prediction model, we measure the correlation between the distributions of the predicted boxes and the ground truth boxes.
In particular, we represent the distribution of bounding boxes as two 2D histograms:
A position histogram for the $(x_\mathrm{stand}, y_\mathrm{stand})$ coordinates, and a size histogram for the $(w, h)$ sizes.
The bin sizes we use for histograms are $15 \times 15$, the same as for the prediction model. 

For this experiment, the ground truth bounding boxes we use are from the validation split of the MS-COCO dataset, which are held out from the training stage. The generated boxes are predicted from the same set of images but with the person segments erased and inpainted. To measure the histogram correlation, we use the  metric:
\begin{eqnarray} 
\label{eq:correlation}
    d(A, B) &=& \frac{\sum_{i}(A_i - \overline{A}) (B_i - \overline{B})}{\sqrt{\sum_{i}(A_i - \overline{A})^2 \sum_{i}(B_i - \overline{B})^2}}
\end{eqnarray}
\noindent
where $A$ and $B$ represent the histograms of the ground truth and the predictions respectively, $\overline{A}$ and $\overline{B}$ are means of $A$ and $B$, and N is the bin count, which is 225 in our case.

Under this proposed metric, the correlation between the ground truth and the prediction is 0.9458 for the position histograms, and 0.9378 for the size histograms. 
As judged by these high correlation scores, our prediction model can mimic real person layouts in natural images. \fig{fig:histogram} shows the 2D histograms we use for this evaluation.

In \fig{fig:heatmap}, we also visualize example heatmaps of the predicted locations (the softmax layer of the location prediction network).
We can see that although our network is trained to predict a fixed unique location, it can approximate the location distribution reasonably well. We include additional heatmaps  of predicted locations in the supplemental.

\begin{figure}[t]
  \centering
  \includegraphics[width=0.48\textwidth]{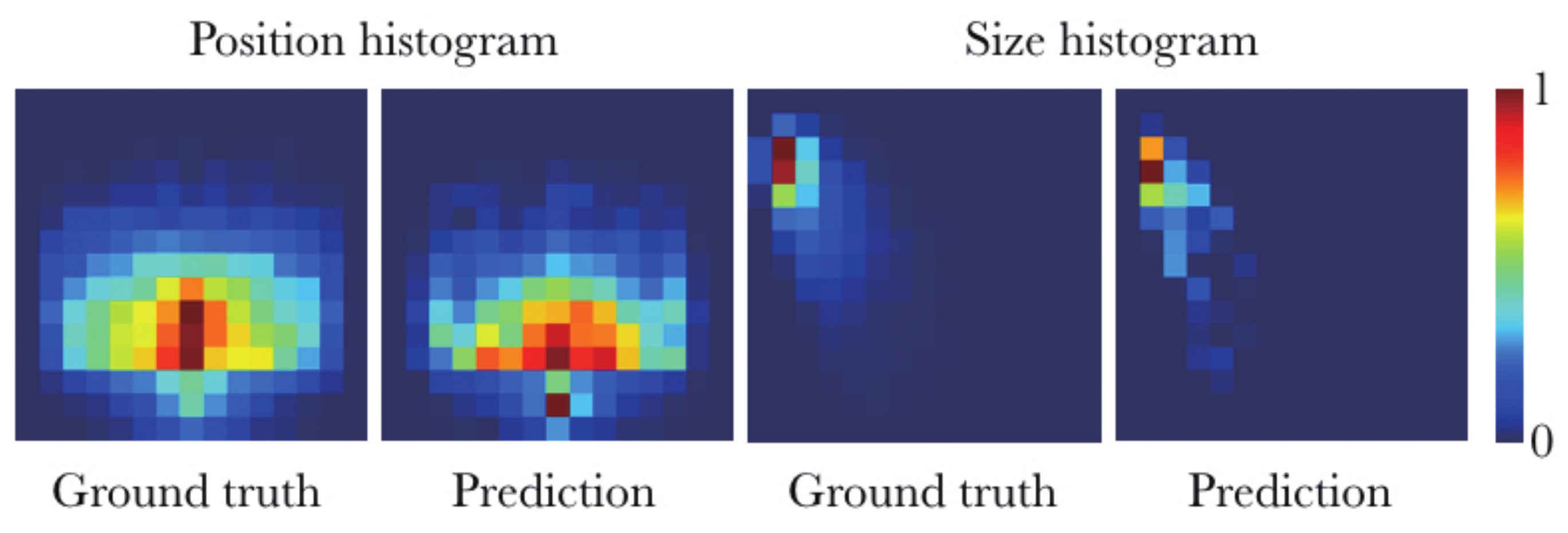}
  \vspace*{-7mm}
  \caption{Ground truth bounding box statistics and statistics measured from our prediction model. For the position distribution, the correlation between  ground truth and prediction is 0.9458. For the size distribution, the correlation between ground truth and prediction is 0.9378.}
  \label{fig:histogram}
  \vspace*{-3mm}
\end{figure}

\begin{figure}[t]
  \centering
  \includegraphics[width=0.40\textwidth,trim={0 45mm 0 0},clip]{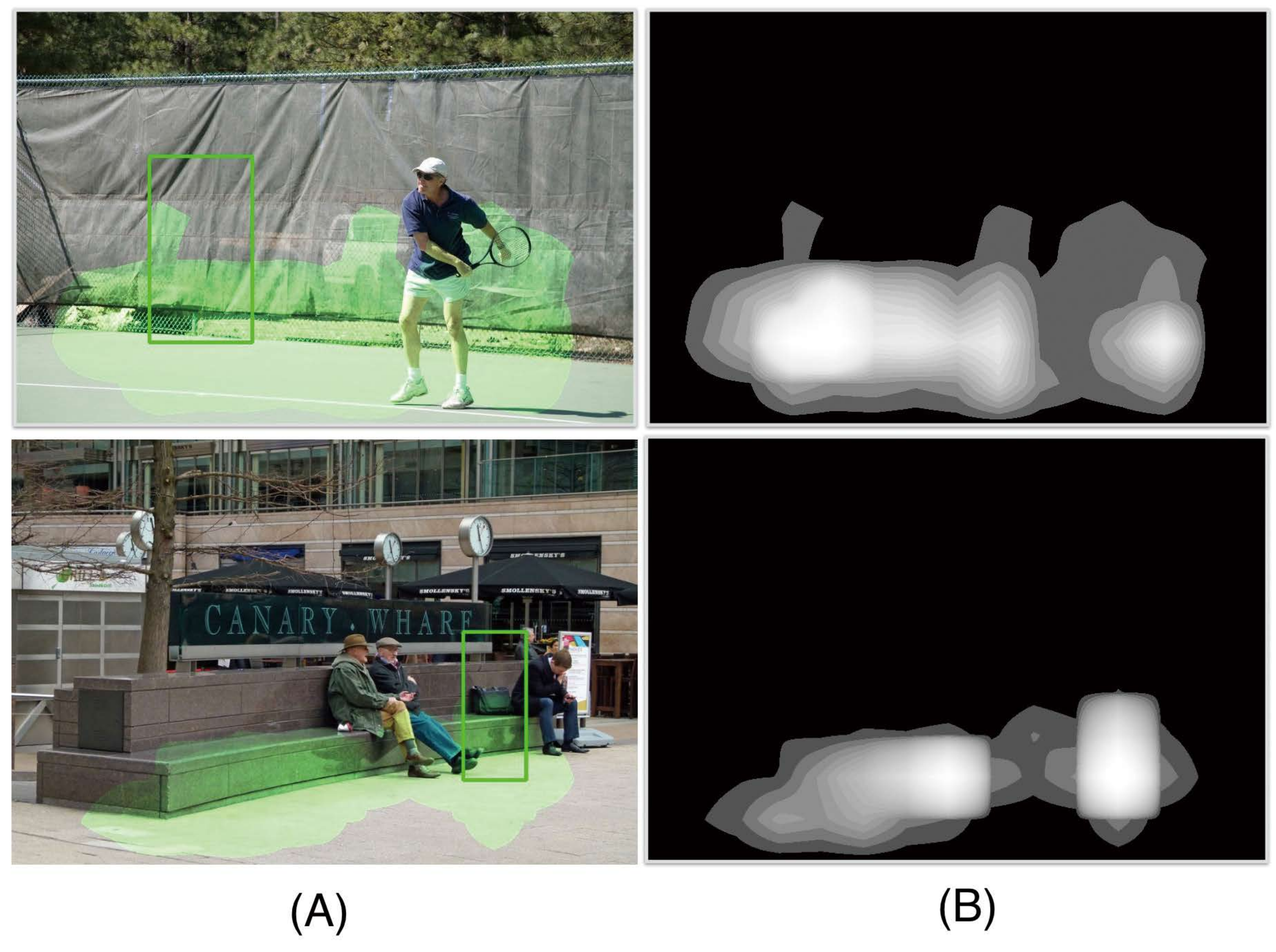}
  \vspace*{-2mm}
  \caption{Example heatmaps of predicted locations. The green boxes show top 1 bounding boxes from our system.}
  \label{fig:heatmap}
  \vspace*{-2mm}
\end{figure}

\begin{table}[t]
    \centering
    \begin{tabular}{L{2.0cm}K{2.5cm}K{2.5cm}}
        \hline Model&Percent real&Percent real\\
        &for textured&for silhouette\\
        \hline 
        Baseline 1&0.176$\pmpad{}$0.054&0.402$\pmpad{}$0.100 \\
        Baseline 2&0.200$\pmpad{}$0.059&0.442$\pmpad{}$0.117 \\
        Baseline 3&0.276$\pmpad{}$0.064&0.505$\pmpad{}$0.109 \\
        Top 1&0.440$\pmpad{}$0.078&0.567$\pmpad{}$0.110 \\
        Best of top 8&0.517$\pmpad{}$0.075&0.742$\pmpad{}$0.107 \\
        Real&0.898$\pmpad{}$0.041&0.864$\pmpad{}$0.075 \\
        \hline
    \end{tabular}\\
    \vspace*{-2mm}
    \caption{Results from the user study. Shown are mean and standard deviations for the percent of images marked real.}
    \vspace*{-4mm}
    \label{tab:user_study}
\end{table}

\begin{figure*}[t]
  \centering
  \includegraphics[width=0.78\textwidth]{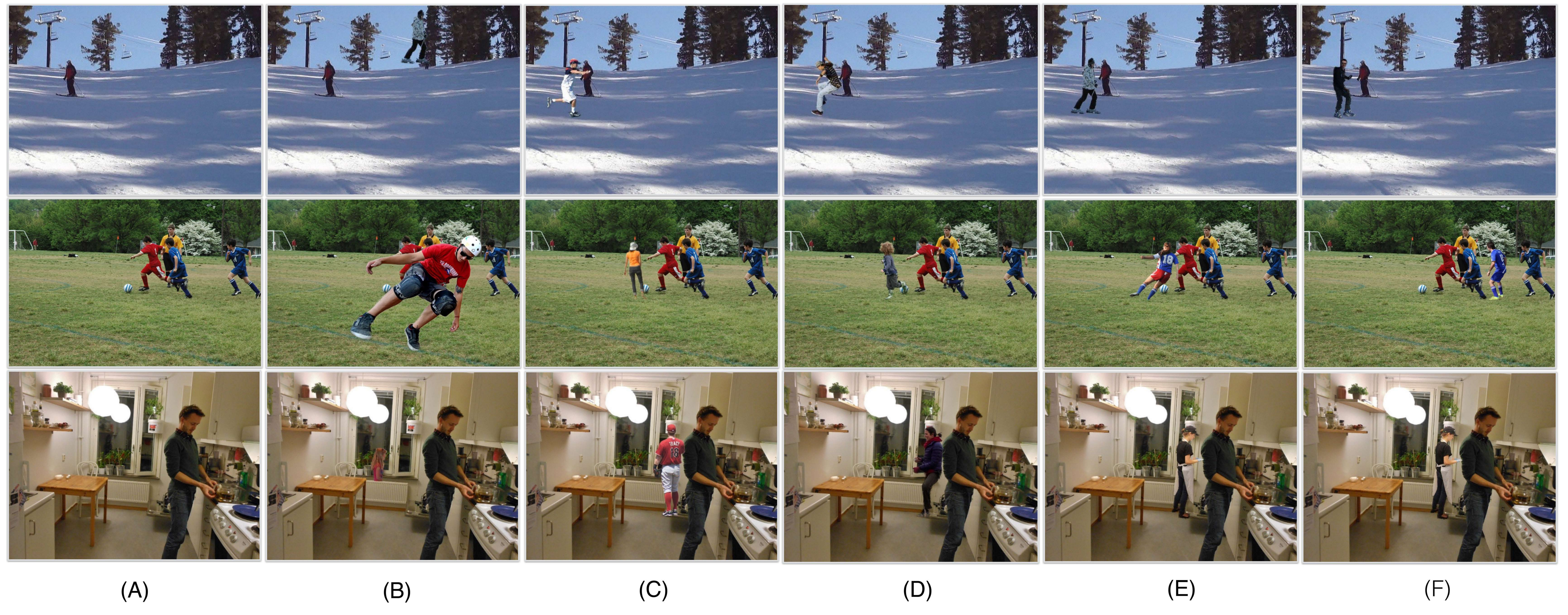}
  \vspace*{-4mm}
  \caption{Examples of the comparison for different methods: (A) input images; (B) baseline 1; (C) baseline 2; (D) baseline 3; (E) top 1; (F) ``best" of top 8.  Note that for the last row, the top 1 composite is also the ``best'' of the top 8 composites.}
  \vspace*{-5mm}
  \label{tab:comparison}
\end{figure*}

\subsection{Qualitative evaluation via user study}
\label{sec:user_study}

To evaluate the visual realism of the composite images, we conduct a human subject study using Amazon Mechanical Turk (AMT). For purposes of comparison, three strong baseline methods are also evaluated, as described below.
\vspace{-0.1cm}
\begin{itemize}
\item \textbf{Baseline 1:} the bounding box is sampled from the ground truth distribution, and the segment is retrieved using our segment retrieval method. This allows us to determine the impact of our bounding box predictions;
\vspace{-0.68cm}
\item \textbf{Baseline 2:} the bounding box is predicted by our system, then a segment is randomly sampled from the candidate pool. 
The purpose of this baseline is to evaluate the impact of our segment retrieval method;
\vspace{-0.25cm}
\item \textbf{Baseline 3:} the bounding box is predicted by our system but the segment is retrieved using a global GIST feature~\cite{Torralba2003b} under Euclidean distance, resembling the work of~\cite{Hays2007}. This allows us to evaluate the effect of deep feature representations on this problem.
\end{itemize}

For our method, we evaluate the top 1 composite from our system.
We also include a manually chosen ``best" images of the top 8 outputs based on the following criteria: combinations of the top 2 location predictions, top 2 size predictions and top 2 retrieved segments. For each background image we evaluate five composite images. Additionally, we believe that future work with larger retrieval datasets or better relighting algorithms could potentially improve results. Thus, to assess the effects of texture and lighting, we also constructed ``silhouette" images where the person's matte is simply filled with a uniform white color.

\subsubsection{User study setup}
During the study, the participants were presented with a sequence of images and told to press R  if an image appears real or F if it appears fake. For the silhouettes we calibrated users by showing them what ``real" silhouettes look like.
For each image, the user had to respond in 10 seconds, otherwise the data was ignored. 
To avoid interference effects, we showed each participant examples from only one method.
For each model, we evaluate 80 composite images. Composite images for different models share the same set of backgrounds.
For quality control, we also included equal numbers of real images and obviously fake composites.
We discarded responses from users who obtained less than 80\% accuracy on the quality control.
For textured images, we collected 25 opinions per image, whereas for the silhouette images, we collected 11 opinions.

\subsubsection{Quantitative results}
Table~\ref{tab:user_study} shows the mean ``realism" scores of each image. 
Standard errors for the scores were computed by applying bootstrapping to the means. 
For the textured images, we notice that both the top 1 and ``best" of top 8 composites outperform all baseline methods.
The ``best" of top 8 composites performs slightly better than top 1.
However, there is still a performance gap between the ``best" of top 8 composites and real images.
One explanation is that our current system has not considered shadows and lighting explicitly.
For the silhouette images, the performances are in the same order but with higher scores of realism.
In particular, the score of the "best of top 8" composites is much closer to that of the real images. This indicates that texture and lighting cues are more frequently responsible for ``giving away" that a composite is not real, as opposed to location, size, and silhouette cues, which give results similar to real images.

For the mean scores, we also tested for significance using a two-sided Student's t-test.
The Holm-Bonferroni method was adopted to control the familywise error rate at the significance level of 0.05.
For the textured images, our method is significantly better than the baselines ($p < 0.0002$).
However, top 1 and ``best" of top 8 are not significantly different ($p = 0.0972$).
For the silhouette images, the best of top 8 method is significantly better than the baseline methods and top 1 ($p < 0.00005$). Top 1 is significantly better than baselines 1 and 2 ($p < 0.0076$), but not significantly different than baseline 3 ($p = 0.2$). For textured and silhouette, real is significantly better than all other models ($p < 0.00002$).



\section{Prototype user interface}
We have also developed a proof-of-concept user interface for composite image generation and interactive layout refinement. Existing compositing tools typically require intensive user interactions, such as finding compatible foreground and background pairs, and finding suitable locations and sizes for composition. We allow users to create and refine composite images with automatic guidance and less manual searching for segments, positions, and sizes.

In our current interface, given an input image, the user is asked how many people should be composited on the background.
The top 1 automatic composite is then returned. 
For each predicted bounding box, 9 candidate segments are also displayed.
The user can then refine the composite by replacing, translating or scaling each person segment. 
\fig{fig:application} shows example results of automatic compositing and user refinement.
Please refer to the supplementary video for more such examples. 

\begin{figure}[t]
  \centering
  \includegraphics[width=0.48\textwidth]{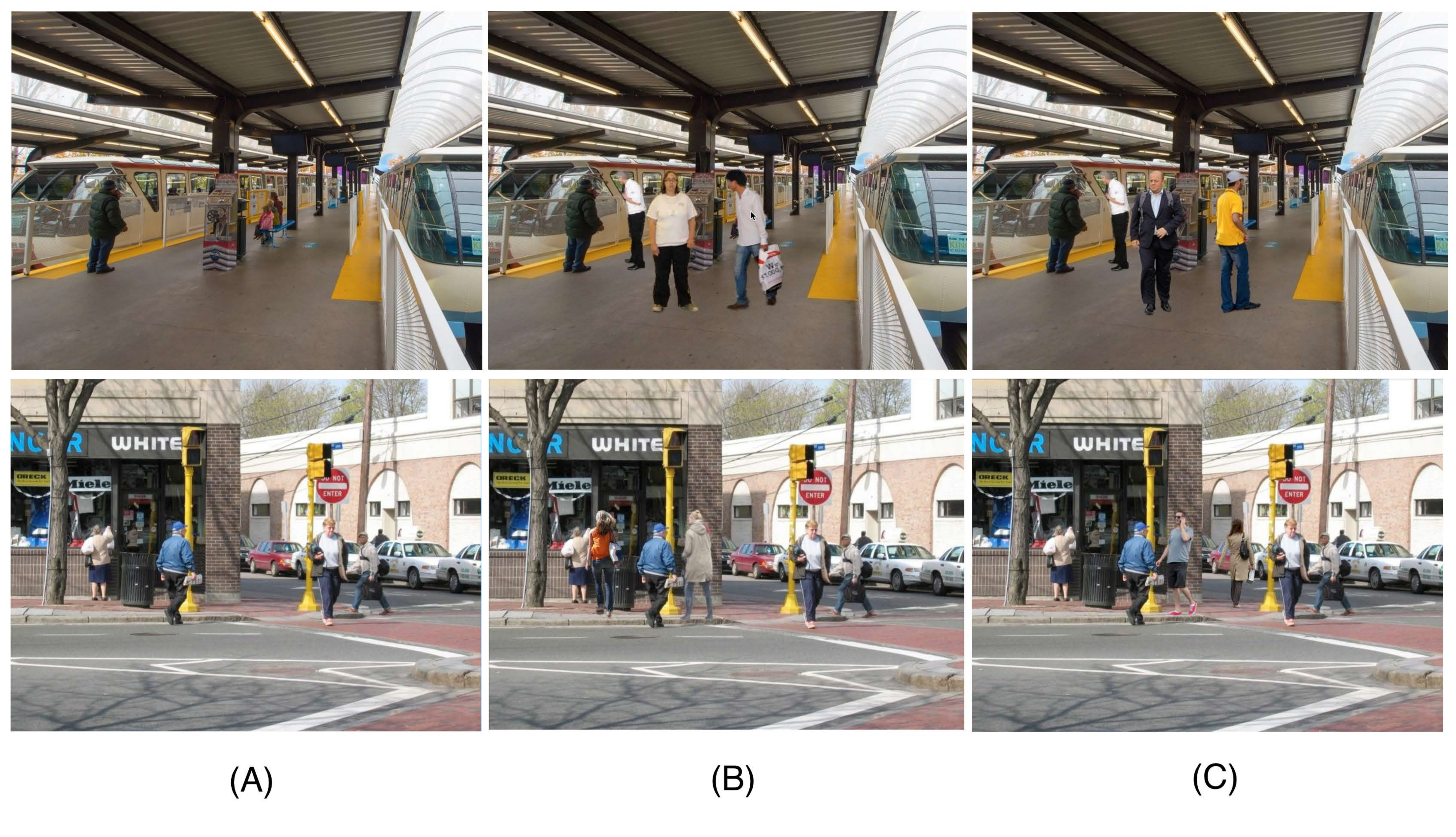}
  \vspace*{-7mm}
  \caption{In our proof-of-concept user interface, given the input image in column (A), users are asked how many people to add to the scene. Column (B) shows  results of our automatic compositing. Column (C) shows the refinement results of (B) via user interactions.}
  \label{fig:application}
  \vspace*{-5mm}
\end{figure}

\section{Limitations and conclusion}
Our current compositing system, however effective, still has a few limitations. 
(1) Although there is no underlying assumption, the outputs from our system tends to bias towards similar positions (e.g. the central region of the image) with similar focal lengths. As all our training and testing images are from standard datasets, we conjecture that the similarity is from the datasets, as people tend to appear in the central regions in natural images.
(2) Our bounding box prediction model depends on the performance of an object detector. There are situations where the results of the detection may hinder the predictions of our model.
(3) While combining the global and local contexts helps retrieving segments that are compatible with the background, the retrieved segments still may not ``interact" correctly with each object in the scene.
\fig{fig:limitation} shows two examples when such interactions are important. 
(4) Our system may potentially benefit from recent success of Generative Adversarial Nets~\cite{Ian2016} by building an end-to-end matting system with adversarial loss.
(5) Our system has not explicitly integrated lighting and shadow consistency with the background. Global relighting methods such as~\cite{Tsai2017} may further improve photo-realism.
(6) Our system has not considered categories other than person and is  not trained end-to-end.


In conclusion, we propose a fully automatic system for semantic-aware person composition.  The system accomplishes compositing by first predicting the bounding box of the potential instance and then retrieving a segment that appears compatible with the local context and global scene appearance. Quantitative and qualitative evaluations show that our system predicts person layouts for a given background scene and outperforms robust baselines.

\begin{figure}[t]
  \centering
  \includegraphics[width=0.35\textwidth,trim={0 45mm 0 0},clip]{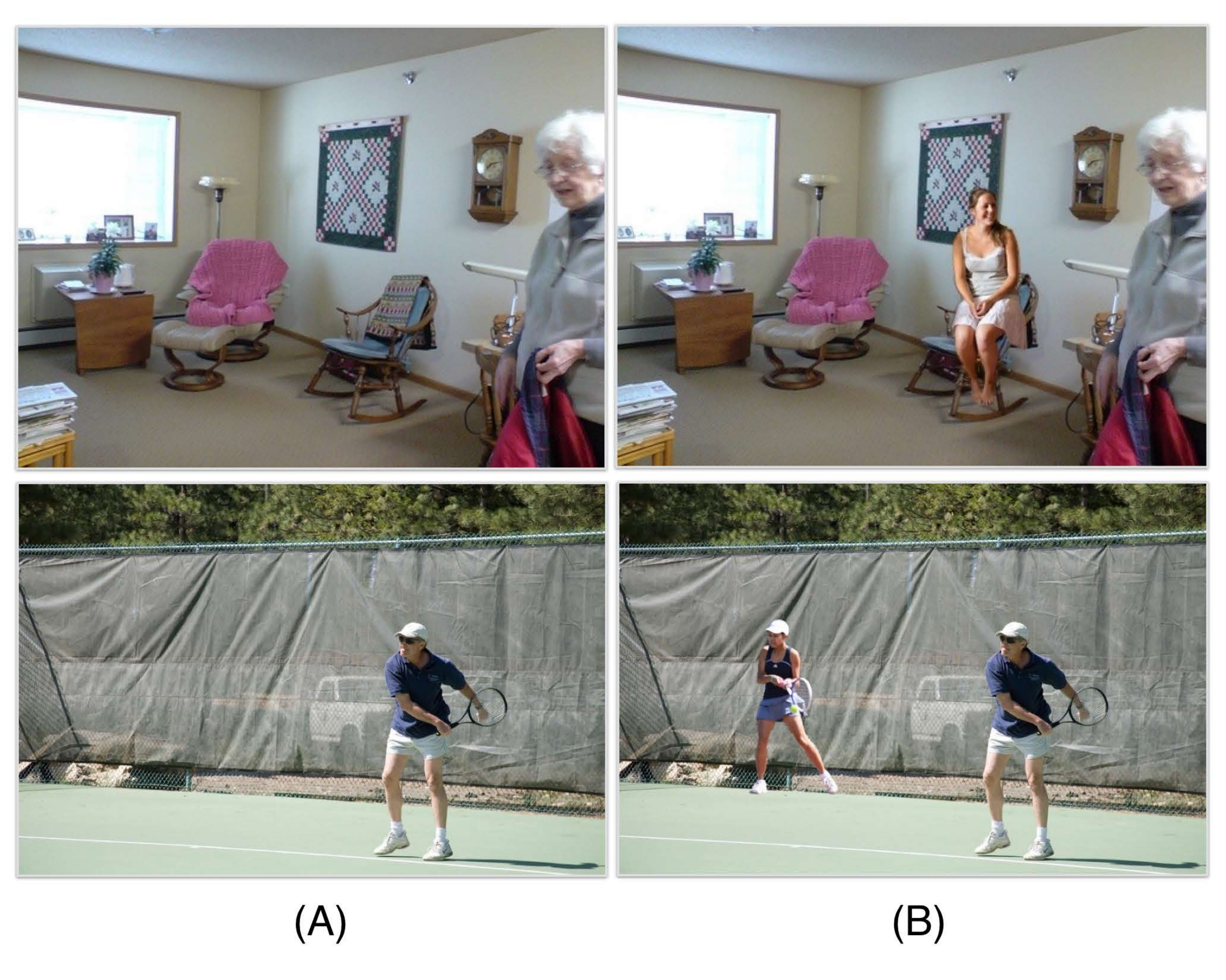}
  \vspace*{-3mm}
  \caption{Limitations of our system. In the first example (top row), although our system retrieves a ``sitting" person segment, it does not align well with the background chair. In the second example (bottom row), the system correctly retrieves a tennis player, but the action is wrong.}
  \label{fig:limitation}
  \vspace*{-5mm}
\end{figure}

{\small
\bibliographystyle{ieee}
\bibliography{egbib}
}

\end{document}